\documentclass[twocolumn,superscriptaddress,footinbib,showpacs,floatfix]{revtex4-1}

\usepackage{graphicx}
\usepackage{dcolumn}
\usepackage{bm}
\usepackage[usenames,dvipsnames]{xcolor}
\usepackage[colorlinks=true,urlcolor=blue,citecolor=blue, linkcolor = blue]{hyperref}
\usepackage{amssymb,amsmath}
\usepackage[percent]{overpic}
\usepackage{cancel}
\def\be{\begin{equation}}
\def\ee{\end{equation}}
\def\ba{\begin{array}{lll}}
\def\ea{\end{array}}
\def\ber{\begin{eqnarray}}
\def\eer{\end{eqnarray}}
\begin{document}
\title{Taming a non-convex landscape with dynamical long-range order: memcomputing Ising benchmarks}
\author{Forrest Sheldon}
\email{fsheldon@physics.ucsd.edu}
\affiliation{Department of Physics, University of California San Diego, La Jolla, CA 92093, USA}
\author{Fabio L. Traversa}
\email{ftraversa@memcpu.com}
\affiliation{MemComputing, Inc., San Diego, CA 92037, USA}
\author{Massimiliano Di Ventra}
\email{diventra@physics.ucsd.edu}
\affiliation{Department of Physics, University of California San Diego, La Jolla, CA 92093, USA}

\begin{abstract}
Recent work on quantum annealing has emphasized the role of collective behavior in solving optimization problems.  By enabling transitions of clusters of variables, such solvers are able to navigate their state space and locate solutions more efficiently despite having only local connections between elements.  However, collective behavior is not exclusive to quantum annealers, and classical solvers that display collective dynamics should also possess an advantage in navigating a non-convex landscape.  Here, we give evidence that a benchmark derived from quantum annealing studies is solvable in polynomial time using digital memcomputing machines, which utilize a collection of dynamical components with memory to represent the structure of the underlying optimization problem.  To illustrate the role of memory and clarify the structure of these solvers we propose a simple model of these machines that demonstrates the emergence of long-range order. This model, when applied to finding the ground state of the Ising frustrated-loop benchmarks, undergoes a transient phase of avalanches which can span the entire lattice and demonstrates a connection between long-range behavior and their probability of success. These results establish the advantages of computational approaches based on collective dynamics of continuous dynamical systems.
\end{abstract}
%
%

%
\maketitle

\section{Introduction}

Non-convex optimization problems draw their difficulty from the complexity of their associated landscapes~\cite{MooreMertens_book}.  These landscapes are often highly corrugated, dotted with hills, valleys and saddles of varying heights which obscure the search for a lowest (or highest) point.  The complexity of this space, combined with the `curse of dimensionality' yields an exponentially large number of potential solutions which are very difficult to prune down by any systematic method.  The innate difficulty and variety displayed by optimization problems, as well as their widespread applications have made their study a continuously active field of research across science and mathematics~\cite{Optimization_book,Optimization_book_intro}.

The exponential growth of the state space with problem size often renders any exact algorithm for locating the optimum impractical as they require an exponential amount of time to sift through the states.  As a result, practitioners must rely on \emph{incomplete} or \emph{approximate} methods which will often generate better solutions in a limited time but are not guaranteed to converge to the exact solution~\cite{kautz2009incomplete,gomes2008satisfiability}.  Despite this, incomplete methods can often converge to the global solution in times orders of magnitude faster than complete solvers~\cite{MAXSAT_competition}.

Early work on approximate methods relied on analogies with the dynamics of physical systems~\cite{kirkpatrick1983optimization} which will minimize their energy as they cool, i.e., during annealing.  For example, to find the ground state of the Ising spin glass~\cite{fischer1993spin},
\begin{equation}\label{eqn:Ising}
E = -\sum_{\langle ij\rangle} J_{ij} s_i s_j , \;\; s_i \in \{-1, 1\},
\end{equation}
simulated annealing gradually improves an initial state $\{s_i\}_{i=1}^N$ by stochastically exploring the state space and steadily lowering an effective temperature~\cite{cocco2006}.  The early success of this approach on combinatorial optimization problems has led to the proliferation of solvers based on a similar stochastic local search and their many variants ~\cite{selman1992new,schoning1999}. Cross pollination with physics has continued, spawning methods such as parallel tempering~\cite{Wang2015}, and quantum simulated annealing~\cite{Santoro2427} as well as the analytical characterizations of combinatorial problems~\cite{mezard2002analytic} and random energy surfaces~\cite{bray2007statistics}.

Annealing has again jumped to the forefront of modern research in the form of quantum annealing and the machines manufactured by DWave~\cite{hen2015probing,DenchevGoogle,Katzgraber2015}.  These machines contain 2-state quantum mechanical elements coupled together in a graph realizing a particular energy function.  During their relaxation, the quantum dynamics of the system allows for collective tunneling of elements through high, thin barriers in the energy function, which may provide some advantage in the search for the optimum.

Similar ideas in the context of cellular automata, neural networks and neuroscience have 
received steady interest~\cite{langton1990computation,chialvo2010emergent}. These examples substantiate the idea that {\it collective behavior} would offer an advantage in the convergence of a solver by allowing for a more efficient exploration of the state space. We then expect that classical solvers which incorporate this feature in their dynamics will have an advantage in both the quality of approximate solutions they produce, and their rate of convergence. 

We also note that annealers admit cluster flipping variants which can drastically increase the rate of thermalization and thus provide faster convergence to the ground state.  Incorporating proposed cluster flips into an annealer as in the Swendsen-Wang and Wolff algorithms can allow the system to overcome larger energy barriers and reduce correlation times~\cite{SwendsenWang,Wolff}.  Modern parallel tempering variants incorporating iso-energetic and thus rejection free cluster moves~\cite{houdayer,zhu} have proven effective at converging to the ground state of spin-glass instances.  While here we have chosen to focus on introducing DLRO in continuous systems, a more detailed comparison between the properties of clusters generated by continuous and discrete solvers and their effect on convergence time is a direction of future work.

The purpose of this work is to explore the presence and advantages of collective dynamics in the context of specific deterministic dynamical systems: Digital memcomputing machines (DMMs)~\cite{UMM,DMM2,DMMperspective}. In DMMs, a combinatorial optimization problem is first transformed into a physical system described by differential equations whose equilibrium points correspond to solutions of the original problem. Theoretical work~\cite{topo,topo1} and simulations of DMMs~\cite{DMM2,DMMperspective,exponential2017speedup,AcceleratingDL} have indicated the presence of long-range order in their dynamics. However, as their native problem form involves several distinct dynamical elements, the complexity of the resulting solver obscures the physical principles underlying its design and function.

Here, we first show that this collective behavior, in the form of {\it dynamical long-range order} (DLRO), 
allows the efficient solution of a class of benchmarks based on the Ising spin glass~(\ref{eqn:Ising}).  Then, by drawing on the structure of the equations governing a DMM, we propose a simplified model that captures several features of their dynamics and illuminates the essential roles played by continuity and memory. 

The paper is organized as follows. In Sec.~\ref{DMMsec}, we briefly introduce the concept of DMMs. In Sec.~\ref{benchmark} we 
discuss the Ising frustrated-loop instances used for benchmarking several methods and show the results of this benchmark. In Sec.~\ref{rigidity} we introduce a simplified model of DMMs that captures its main features. In Sec.~\ref{DLROsec} we  discuss the dynamical long-range order that emerges in these dynamical systems. In Sec.~\ref{conclusions} we offer our conclusions.

\section{Digital Memcomputing Machines} \label{DMMsec}

In this section we present a very brief introduction to the concept of DMMs~\cite{DMM2}. A thorough discussion of the physics behind them 
and the problems they have been applied to can be found in the perspective article~\cite{DMMperspective}.

DMMs are dynamical systems designed as circuit elements to solve circuit satisfiability (SAT) problems~\cite{UMM,DMM2,DMMperspective}, 
so that despite operating in continuous time, initial and final states are digital, hence the machines are scalable. A particular problem may be translated into circuit SAT format as a combination of AND, OR, and NOT gates which are then replaced with dynamical circuit elements.  The elements are composed of resistors, capacitors, voltage/current generators and resistors with memory (memristors) whose dynamics conspire to lead the circuit voltages to a state where all logical constraints are satisfied.

These elements are governed by ordinary differential equations (ODEs) and constraints imposed by Kirchoff's laws and thus they may be efficiently simulated.  
Numerical studies have shown that DMMs are effective at solving a wide range of combinatorial optimization problems~\cite{exponential2017speedup,DMMperspective,AcceleratingDL,ILP}. For instance, DMMs have proven effective at solving Integer-Linear Programming problems  (a DMM solver compared favorably to standard algorithms when applied to instances in the MIPLIB 2010 library and was able to establish the feasibility of an instance for which this was previously unknown~\cite{ILP}), and maximum satisfiability (MAXSAT) problems derived from XORSAT where DMMs displayed linear scaling in their approach to a satisfiable threshold while all other solvers tested scaled exponentially~\cite{exponential2017speedup}. 

\section{Frustrated-Loop Instance Benchmarking}\label{benchmark}

Recent progress on fabricating quantum mechanical hardware has generated renewed interest in problems of the form~(\ref{eqn:Ising}), or as it is known in the optimization community, quadratic unconstrained binary optimization (QUBO), which can be mapped into a MAXSAT problem~\cite{computational_complexity_book}.  When the couplings $J_{ij}$ are assigned randomly, this is known as the Ising spin-glass and is a simplified model for the behavior of glassy systems~\cite{fischer1993spin}.  Studies of the thermodynamic properties of these systems have shown that, below a certain temperature, the phase space  can become separated into clusters of states and may admit an exponential number of metastable states.  This proliferation of metastable states obscures the search for the global minimum and gives a physical interpretation for the computational hardness of the optimization problem.

The problem of benchmarking MAXSAT solvers is generally hindered by the fact that the problems are NP-hard, and, for an arbitrary instance even confirming a solution may require exponential time~\cite{barahona1982computational}.  For this reason, planted solution instances are commonly employed in which instances are generated such that they have a known solution~\cite{jia2004}.

\begin{figure}
	\begin{center}
		\includegraphics[width=8.6cm]{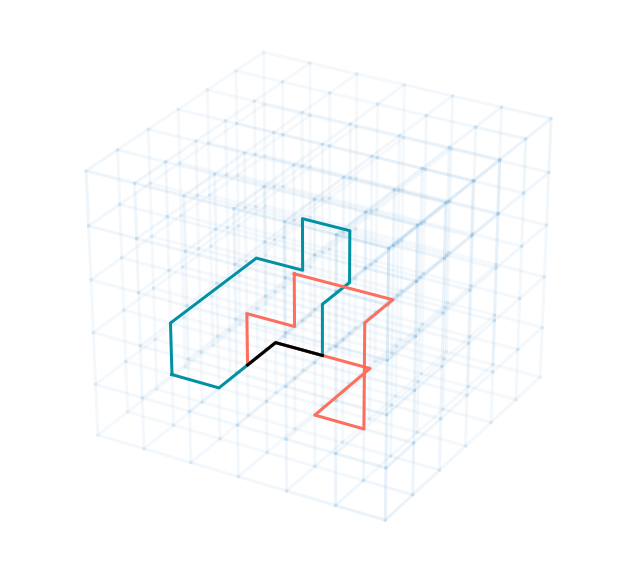}
		\caption{A schematic representation of instance creation. Separate frustrated loops (blue and red curves) are generated by random walking around the lattice until the walk crosses itself.  Each loop has its own Hamiltonian consisting of $J_{ij} = 1$ for all bonds except one with $J_{kl} = -1$ such that the ground state of the loop will have one unsatisfied bond.  When the loops are combined, overlapping bonds (shown in black) have a coupling $J_{ij}$ which is the sum of the contributions from each separate loop.
		\label{fig:loops}}
	\end{center}
\end{figure}

Benchmarking studies on quantum annealing have introduced the class of Frustrated-Loop Hamiltonians~\cite{hen2015probing,king2015performance} in which the total Hamiltonian is written as the sum of Hamiltonians of a set of loops containing a single frustrated bond (see schematic in Fig.~\ref{fig:loops}),
\begin{equation}\label{eqn:frustham}
H = \sum_i H_{FL,i}.
\end{equation}
The loops are formed such that the planted solution minimizes all of the Hamiltonians $H_{FL,i}$ simultaneously, and so minimizes their sum.

In order to generate these instances, we first construct an underlying lattice which we take to be hypercubic in $D$-dimensions with periodic boundary conditions.  Each loop is generated by beginning at a randomly selected site and performing a random walk until it crosses itself.  The length, $l$, of the loop formed is generally required to be above some limit, otherwise it is rejected.  For example, the instances solved on DWave employ a loop length limit of 
$l\ge 8$~\cite{hen2015probing}. It is also noted that discarding the length limit seems to lead to very difficult instances, although an explanation for this feature is not given.  In our investigations of the instances, we found that discarding the loop length limit leads to instances of widely varying difficulty, and that both the variance and mean of the solution time (measured with simulated annealing) decreased as the length limit increased.  In order to avoid the complications of a widely varying difficulty, while generating the most difficult available instances, we then chose a length limit of $l\ge 6$ for our generated instances.

In order to generate a loop, we consider planting the ferromagnetic solution $s_i=1$.  After generating an instance, any other solution may be hidden by means of a gauge transformation.  All interactions in the loop are chosen to be ferromagnetic, $J_{ij}=1$, except one which is selected at random to be anti-ferromagnetic $J_{ij}=-1$.  The solution to the loop hamiltonian $H_{FL,i} = -\sum_{\langle ij\rangle \in l_i} J_{ij} s_i s_j$ is thus an assignment with one unsatisfied interaction.

The number of loops, $M$, generated must be proportional to the number of sites $N=L^D$ and may be characterized by a \emph{density} $\alpha$ such that  $M=\alpha N$.  These instances are known to demonstrate a \emph{hardness peak} in $\alpha$ such that the most difficult instances are generated when there are neither too few loops, in which case they do not overlap and each may be solved separately, nor too many, in which case the antiferromagnetic interactions tend to be canceled by the more numerous ferromagnetic interactions~\cite{hen2015probing,albash2018demonstration}.  The value of $\alpha$ at the peak also tends to align with the amount of frustration in the instance, as measured by the number of unsatisfied interactions in the ground state.

In order to generate difficult instances, in $D=2$ dimensions we used a simulated annealing solver to test instances across a range of $\alpha$, finding that the most difficult instances lay at $\alpha \approx 0.2$, consistent with the results on the pseudoplanar chimera graphs in~\cite{hen2015probing}.  For $D=3$ dimensions, the optimal value of $\alpha$ was estimated using the amount of frustration in the instances  as suggested in~\cite{hen2015probing} and found to lie at $\alpha\approx 0.3$.

Benchmarking was carried out using instances generated on a 3-dimensional hypercubic lattice with periodic boundaries . The implementation of the dynamical equations of DMMs as in Ref.~\cite{DMM2} was appropriately modified to handle the Ising frustrated loop instances expressed as a maximum satisfiability problem in conjunctive normal form~\cite{complexity_bible} (see the Supplemental Material below for a discussion of this transformation). These were then simulated using a commercial sequential MATLAB solver dubbed Falcon provided by MemComputing, Inc. In addition, we have implemented two standard annealing algorithms in Python (simulated annealing (SA) and parallel tempering (PT)), as well as used a well-known commercial mixed-integer programming solver, IBM CPlex~\cite{cplex}. Since Falcon was implemented in interpreted MATLAB and the focus was on scaling rather than runtime, we used only the simplest implementation of each solver but performed substantial tuning. Details of the implementation and tuning on the instance class for SA and PT, as well as the configuration for IBM CPlex can be found in the Supplemental Material.  

All solvers were run on frustrated-loop instances in 3 dimensions, ranging in size from $L=6$ (total number of spins $N=216$) to $L=40$ ($N=64,000$).  The sizes used for tuning were included for the annealers (SA and PT) while CPlex and Falcon were run on sizes $L\ge 10$.  Comparisons between solvers are hampered by ambiguity in the efficiency of the implementation and differences in processor speed when run on different machines.  In order to skirt the second of these ambiguities we have displayed the solutions in terms of estimated floating point operations  (flop's) calculated by multiplying solution time by the peak floating point operations rate of a single core.  However we emphasize that when comparing different implementations only differences in \emph{scaling} are relevant.

As is clearly visible from Fig.~\ref{fig:scaling}, the memcomputing solver converged to the exact ground state with superior scaling to all solvers tested, allowing us to achieve sizes much larger than possible with other solvers within the time limitation of $\approx 10^6$ sec.  The total number of flop's appears to scale approximately as $ N^{1.5}$ at large sizes, while all other solvers appear to scale exponentially as $ \exp(b N^c)$, with $b$ and $c$ solver-specific constants reported in the Supplemental Material and shown in Fig.~\ref{fig:scaling}. Details of the fitting procedure  as well as the figure displaying the recorded time to solution may be found in the Supplemental Material.

\begin{figure}[t]
	\begin{center}
		\includegraphics[width=8.6cm]{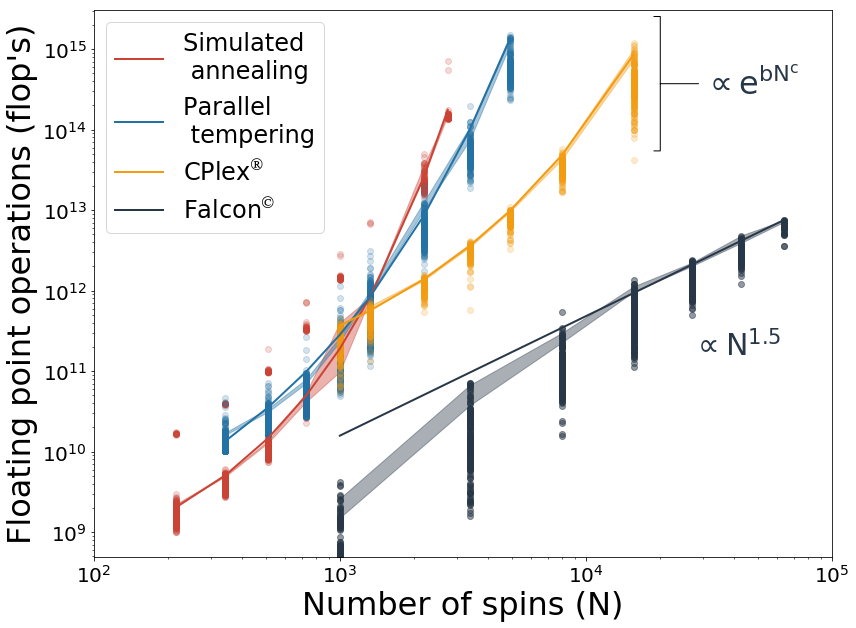}
		\caption{ Scaling of floating point operations necessary for different solvers to reach the ground state of the 3D frustrated-loop instances as a function of the total 
			number of spins $N$. Solution times have been converted to estimated number of floating point operations. The sequential memcomputing solver implemented in MATLAB is dubbed Falcon and was run on an Intel Xeon 6148. Varying numbers of instances were run at each size and solver depending on required computation time (See the Supplemental Material for details). Comparisons with simulated annealing (SA), parallel tempering (PT), and IBM CPlex  run on an Intel Xeon E5430 are also shown. All calculations were performed on a single core. The solid lines are the best fits of the 95th quantile time to solution for all four solvers. The exponential fits have the following parameters: for IBM CPlex, $b=0.12$ and $c=0.46$, for SA, $b=0.069$ and $c=0.67$, and for PT $b=0.32$ and $c=0.46$.}\label{fig:scaling}
	\end{center}
\end{figure}

\section{A Simplified DMM Model}\label{rigidity}

As mentioned in Sec.~\ref{DMMsec}, a DMM is constructed in correspondence to the logical circuit it will solve. For example, the subset-sum problem studied in~\cite{DMM2} utilizes a circuit with the same structure as one used to add a subset from a group of numbers. Each traditional logic gate is replaced by a \emph{self-organizing logic gate} consisting of a set of interconnected input and output terminals, each of which is dressed with a number of memristors (resistors with memory), resistors, capacitors, and voltage/current generators forming a \emph{dynamic correction module} (DCM)~\cite{DMM2}.  When voltages are applied to the boundaries of the circuit, the dynamics of these elements are configured to satisfy the constraints enforced by each gate, and lead the circuit to a state where no logical contradictions are present.

However, the restrictions imposed by the native hardware formulation of DMMs necessarily complicate their design and obscure important features of their implementation.  Moreover the dependence on a large number of physical parameters in these models makes understanding their dependence on these parameters difficult.  As such, an effective model that discards the constraints of hardware and that is more tractable for analysis is desirable.  In the following we formulate a simple model that replicates several features of the full implementation of digital memcomputing machines and we use this to probe the existence of DLRO in the search for a solution.  

For our purposes, finding the ground state of the Ising system provides the advantage that it can be expressed in terms of very simple homogeneous constraints leading to a concise set of  equations. In addition, its real-space lattice representation allows for a clearer demonstration of DLRO since the real-space distance of the lattice corresponds to the distance in the constraint graph. 

\begin{figure}
	\begin{center}
		\includegraphics[width=8.6cm]{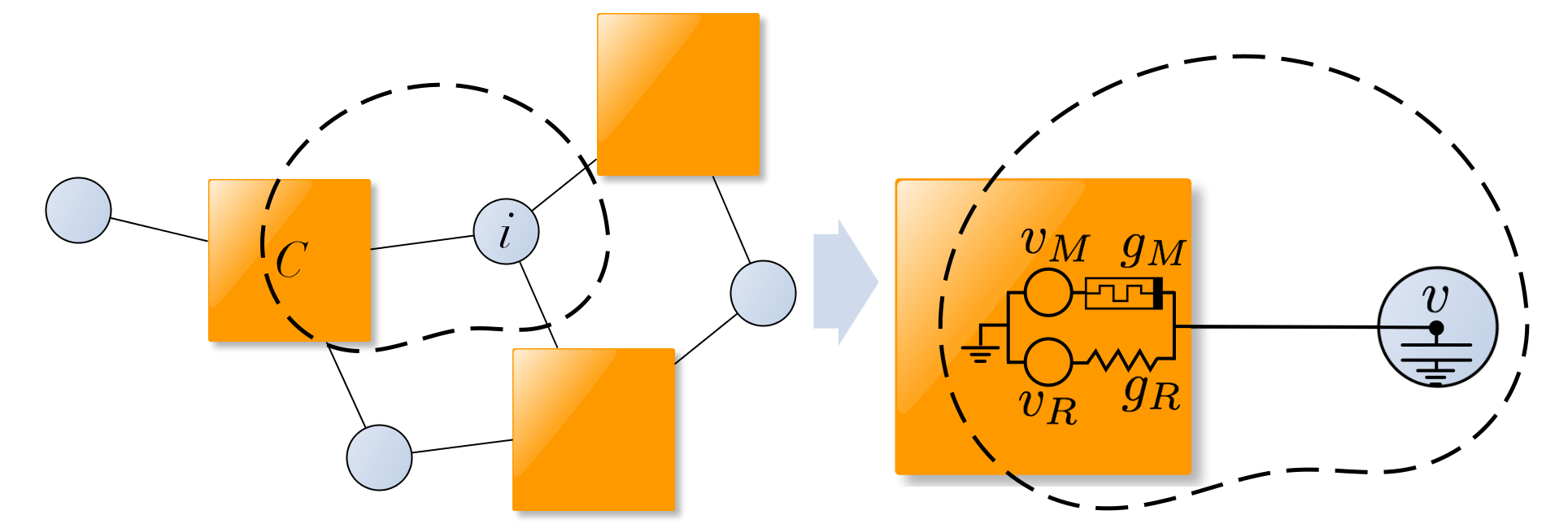}
		\caption{ An arbitrary constraint (C) satisfaction problem expressed as a factor graph can be translated into an electrical circuit with memory by considering the effect of each constraint on the site $i$. $v_{R}$ is a voltage generator and $g_R$ is the conductance of a standard resistor.  $v_{M}$ is a voltage generator and $g_M$ is the conductance of a resistor with memory. 
			\label{fig:factor}}
	\end{center}
\end{figure}

We consider the contribution of constraint $C$ to the dynamics of site $i$ (see Fig.~\ref{fig:factor})~\cite{DMM2}. The dynamics of the circuit are constructed such that the voltage generators impose the logical constraint on the voltage $v_i$ at site $i$.  The memristor conductance $g_M$, sensing a current flowing across it due to an unsatisfied constraint, will alter its value to accelerate the convergence of $v_i$ to the logically-consistent solution.  Generally, this is accomplished by increasing the memristor conductance, thus allowing more current to flow into or out of the site. As memristors are polar objects, complex constraints may require several memristors and generators to accomplish this, accounting for the number of memristors in DCMs~\cite{DMM2}.

A few simplifying assumptions give the general form for the contribution of constraint $C$ to site $i$ as~\cite{DMM2},
\begin{align}
\dot{v}_i & = \Delta g_{M} x \Delta V_{M} + g_{R} \Delta V_{R}, \label{vdot} \\
\dot{x} & = h(\Delta V_{M}, x), \;\;\;\;x\in [0, 1],\label{xdot}
\end{align}
for the voltage $v_i$ representing the variable $i$ and the memory state variable of the memristor $x$.  We can regard the first and second terms on the rhs of Eq.~(\ref{vdot}) as representing the \emph{total} memristive and resistive contributions from the DCM, respectively.  These are weighted by the conductances $\Delta g_M$ and $g_R$, respectively, into which we have absorbed a capacitive timescale.  We regard the memory state variable $x$ and function $h$ in 
Eq.~(\ref{xdot}) as an effective representation of the state and evolution of all memristors in the DCM, giving us considerable freedom in choosing the form of $h$. 

These equations bear a close resemblance to those of Lagrange programming neural networks (LPNNs) proposed in~\cite{zhang1992lagrange,nagamatu1996stability} and the dynamical systems proposed in~\cite{ercsey2011optimization}.  In these works a Lagrangian, $\cal L$, for a constraint satisfaction problem on variables $\{s_i\}$ is formed from a set of constraint functions $C_m (\{s_i\}) $ which vary from 0 when the constraint is satisfied to 1 when unsatisfied and a set of weights for each constraint $x_m$, ${\cal L} = \sum_m x_m C_m (\{s_i \})$.  In the case of LPNNs, the equations of motion of the system are then derived as
\begin{align}
\dot{s_i} &= -\nabla_{s_i}{\cal L} = -\sum_m x_m \nabla_{s_i} C_m, \label{eqn:LPNN} \\
\dot{x}_m &= \nabla_{x_m} {\cal L} = C_m,
\end{align}
which in our formulation (Eqs.~(\ref{vdot}) and~(\ref{xdot})) would correspond to an {\it unbounded}, voltage-controlled set of memristors with equal weight.  In~\cite{ercsey2011optimization} the equations for the multipliers are altered to $\dot{x}_m = x_m C_m$, 
which has the effect of making the system hyperbolic, and is analogous to choosing {\it unbounded} current-controlled memristors in Eq.~(\ref{xdot}).   The dynamics of both systems are such that the variables $s_i$ of the optimization problem act to minimize the energy, while the weights $x_m$ act to increase it, forming a sort of competitive dynamics which seek out saddle points in the Lagrangian. The weights may be re-expressed as an integral memory term in the $s_i$ equations and so may be interpreted as ``memory terms.''  

The continuous constraint weighting that these Lagrangian methods perform bears a close resemblance to DMMs. However, the existence of DLRO in these systems has never been explored.  In the following we propose a system that combines the bounded motion of memristive variables and attempts to explicitly include a form of long-range order in their dynamics.  The model produces a transient phase of avalanches similar to those seen in DMMs~\cite{DMMperspective,topo,topo1,borel} and allows us to continuously control the extent to which both the memory and DLRO are present.  Using this we show that the transitions in these avalanches are capable of flipping an extensive number of variables in the system and that both the correlation length and success probability are modulated by the range of these memory variables.  Most importantly, the behavior this system shows is a consequence of continuity and does not have discrete analogues.

The inspiration for a simplified model of DMMs draws on the notion of \emph{rigidity} in a condensed matter system.  For example, the breaking of translational symmetry also coincides with presence of an elastic potential energy contribution leading to the collective motion of a solid phase~\cite{anderson1984basic}.  The presence of a continuous symmetry in the equations and its effective breakdown can give rise to behavior analogous to zero-modes in statistical physics/field theory~\cite{Book_Peskin}. As a consequence, along some directions of the phase space the system can respond in a correlated, or `rigid' manner in which large clusters of variables will transition together~\cite{topo,topo1}.

For example, in a lattice of continuous ``spins" obeying (here $\sigma'(\cdot)$ is the derivative of $\sigma(\cdot)$, giving a rect function which limits the range of $v_i$ to $[-1, 1]$),
\begin{eqnarray}\label{eqn:rigid}
&\dot{v}_i = -\sum_j |J_{ij}|\big(\sigma(v_i) - \mathrm{sgn}(J_{ij})\sigma(v_j)\big)\sigma'(v_i), \\
&\sigma(x) = \begin{cases}
1, & x > 1\\
x, & -1 \le x \le 1 \\
-1 & -1 < x,
\end{cases}
\end{eqnarray}
the system will exponentially relax to a state in which every variable $v_i$ takes the value $\text{sgn}(J_{ij})v_j$ for all of its neighbors $v_j$.  If the underlying lattice is ferromagnetic ($J_{ij} = 1)$, then taking any spin to its limiting value $v_i = \pm 1$ via an external field will cause the entire lattice to transition with it in a manner analogous to long-range order.  In contrast, for a discrete system in the ferromagnetic state $s_i =  1$, flipping a single spin will not cause the rest of the lattice to transition as it will not change the sign of any local fields. The ability of a local perturbation to flip large clusters of spins might benefit a solver attempting to satisfy a constraint while maintaining the satisfaction of its neighbors.

This may be viewed as gradient motion on the potential,
\begin{align}
E =& \frac{1}{2}\sum_{\langle ij\rangle}|J_{ij}|\big(\sigma(v_i) - \mathrm{sgn}(J_{ij})\sigma(v_j)\big)^2\\
 =& \frac{1}{2} \sigma(\vec{v}) L_J \sigma(\vec{v})
\end{align}
where $L_J$ is a weighted graph Laplacian.  For a satisfiable instance, this operator will always have a zero eigenvalue corresponding to the ground state and which connects the two satisfied states.  However, the presence of unsatisfiable/frustrated constraints renders this impossible in the above model: in this case all spins will relax to $v_i  =0$ and pulling a single spin to $\pm 1$ will not propagate through the lattice.

\begin{figure}[t]
	\begin{center}
		\includegraphics[width=8.6cm]{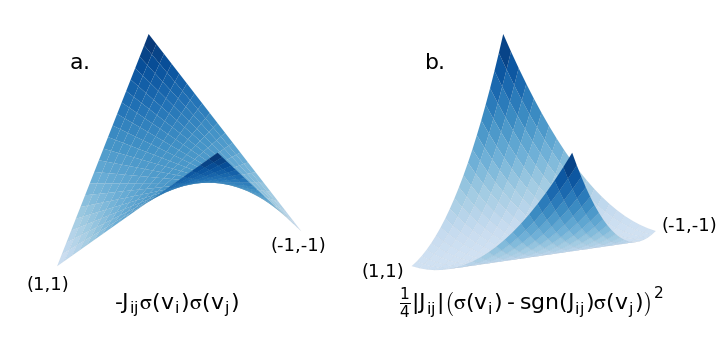}
		\caption{During the dynamics, memristive variables $x_{ij}$ act to interpolate between gradient dynamics on (a) the original interaction, when a constraint is unsatisfied $(x_{ij} \to 1)$, and (b) the `rigid' interaction when the constraint is satisfied $x_{ij} \to 0$.  This removes the energy barrier between the two states that satisfy the interaction, allowing transitions of large clusters of variables.  The satisfied configurations are denoted for the ferromagnetic case, $v_i = v_j$ and below each surface we give the effective energy function on which the voltages follow gradient dynamics.	\label{fig:interaction}}
	\end{center}
\end{figure}

We can combine the weighted gradient dynamics of the LPNN type systems~(\ref{eqn:LPNN}) with terms that induce rigidity~(\ref{eqn:rigid}) by utilizing the bounded motion of memristive variables to smoothly transition between these two interactions,
 \begin{align}
 \dot{v}_i  &= \sum_{j} J_{ij} x_{ij} \sigma(v_j) - \nonumber \\
 &\quad(1-x_{ij})\frac{|J_{ij}|}{2}(\sigma(v_i) - \text{sgn}(J_{ij})\sigma(v_j)), \quad v_i \in [-1, 1],\label{vdotrigid} \\
 \dot{x}_{ij} & = \beta x_{ij}(1-x_{ij}) \Big(|J_{ij}|\big(1-\text{sgn}(J_{ij}) v_i v_j\big) - \gamma\Big).\label{xdotrigid}
 \end{align}
When $x_{ij}\approx 1$ the voltages follow the fields imposed by the neighboring voltages as in an LPNN with ${\cal L} = \sum_{\langle ij \rangle} x_{ij} |J_{ij}|(1-\text{sgn}(J_{ij}) v_i v_j)$, causing them to take the integral values $v_{i} = \pm 1$, which agree with the sign of the overall local field.  Constraints which are satisfied by this configuration will then see their contribution to the field reduced as $x_{ij} \to 0$. These constraints then move under the equation of motion $\dot{v}_i = -\frac{|J_{ij}|}{2}(\sigma(v_i) - \text{sgn}(J_{ij})\sigma(v_j))$ which allows $v_i$ and $v_j$ to transition collectively between the two satisfied states of the constraint. The two interaction terms interpolated between are displayed in Fig.~(\ref{fig:interaction}). The voltages $v_i$ are limited to the interval $[-1, 1]$. From a state of the dynamical system, the spins of the original Ising model~(\ref{eqn:Ising}) are assigned as $s_i = \mathrm{sgn}(v_i)$ such that the spins of the Ising model undergo the orthant dynamics of the underlying continuous voltages. 

The memory state follows the simplest equation for a {\it bounded}, volatile memristor subject to an effective voltage $|J_{ij}|(1-\text{sgn}(J_{ij}) v_i v_j)$.  This voltage is the energy with which the constraint is violated, and the constant $\gamma$ sets a threshold below which $x_{ij}$ will begin to decay.  The constant $\beta$ indicates that the memristive timescale is generally different from the voltage timescale (set by the $RC$ constant at the node) which will play an important role in our analysis of the system.  While the memory variables do not directly interact, their coupling through the voltages leads to an effective interaction.  For the memristive network model shown in~\cite{caravelli2017complex,caravelli2018mean} it was shown that the network topology leads to pairwise interactions obeying a Lyapunov function such that despite the lack of an explicit interaction term in the dynamical equations, interactions between memristive variables are present through coupling with the voltages.

To clarify the dynamics of this model and the function of the rigidity terms, we consider a two-spin system coupled with $J_{12} = 1$, where the first spin is subject to a small local field $1 > h > 0$.  Initializing the system with $x_{12} \approx 1$, the voltages will initially obey
\begin{align}
\dot{v}_1 = \sigma(v_2) + h\\
\dot{v}_2 = \sigma(v_1).
\end{align}
If set near $v_1=v_2 = -1$ the voltages will quickly relax to this state, satisfying the $J_{12}$ constraint and causing $x_{12}$ to decrease.  For a small field $h$, the voltages will remain at $-1, -1$ until $x_{12}\approx 0$ at which point they will obey,
\begin{align}
\dot{v}_1 = -(\sigma(v_1) - \sigma(v_2)) + h\\
\dot{v}_2 = -(\sigma(v_2) - \sigma(v_1)).
\end{align}
The local field $h$ will now cause $v_1$ to drift in the positive direction and the interaction will cause $v_2$ to drift along with it, causing both spins to transition collectively to $v_1 = v_2 = 1$.

Over the course of the dynamics, the ``rigidity terms'' above allow voltages to form clusters with satisfied constraints that are capable of transitioning together under the influence of neighboring unsatisfied constraints. This has a dramatic effect on the dynamics, and inclusion of these ``rigidity terms'' to the gradient-like first terms in Eq.~(\ref{vdotrigid}) act to ensure that these transitions maintain the satisfaction of these clusters.  The form of the interactions that are transitioned between are displayed in Fig.~\ref{fig:interaction} such that after relaxing to one of the two states which satisfy the constraint,  the decay of $x_{ij}\to 0$ removes the barrier between the two states allowing the simultaneous transition of the two variables between them. Any unsatisfiable spin system may be associated with one or several satisfiable instances formed by removing any unsatisfied bonds in the ground state.  The dynamics of the system attempt to discover the underlying satisfiable instance as sub-lattices where $x_{ij} \to 0$.

\begin{figure*}[t!]
	\begin{center}
		\includegraphics[width=17.2cm]{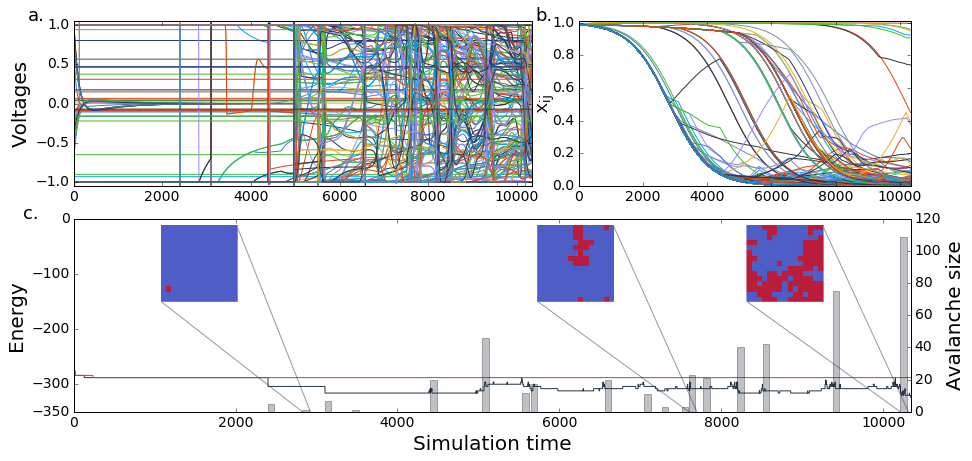}
		\caption{ When the model of Eqs.~(\ref{vdotrigid}) and~(\ref{xdotrigid}) is simulated for a 2-dimensional instance ($L=15$) under a separation of timescales ($\beta=\frac{1}{400}$), the voltage trajectories (a) evolve under a series of sharply defined avalanches due to the slow motion of the memristors (b) modifying the clause weights.  In (c) we have plotted the energy (left axis) without the influence of memristors (red, $\beta=0$) and with them (black, $\beta=\frac{1}{400}$, $\gamma= 0.65$) showing that the motion of the memory variables allows the system to reach a far lower energy, and ultimately the ground state.  The sizes of the avalanches (c, right axis) are plotted as gray bars, showing that their size grows over the course of the simulation until a large avalanche brings the system to its ground state.  The avalanches are depicted in the inset in red with the rightmost inset corresponding to the largest avalanche in the run. 
			\label{fig:trajectories}}
	\end{center}
\end{figure*}

We simulate the system described by Eqs.~(\ref{vdotrigid}) and~(\ref{xdotrigid}) from random initial voltages and $x_{ij}(0) = 0.99$, integrating the equations of motion until the energy~(\ref{eqn:Ising}) (calculated from the signs of the voltages) has reached the planted ground state or some maximum time has elapsed. This is typically chosen quite long, such that the system solves an instance with a probability $p\approx 0.95$ for a given initial condition.  For a more detailed discussion of the numerical implementation, see the Supplemental Material. A typical run, showing the voltages, memristances and energy of the system is shown in Fig.~\ref{fig:trajectories} on a 2-dimensional instance, $L=15$, where we also show that in the absence of constraint weighting via the memory variables ($\dot{x}_{ij}=0$) the system is unable to reach the ground state (the red curve in Fig.~\ref{fig:trajectories}(c)).  In this case the system undergoes gradient dynamics and converges to a local minimum of ${\cal H} = -\sum_{\langle ij\rangle} J_{ij} v_i v_j, \; v_i \in [-1, 1]$.  The action of the memory variables may be interpreted as slowly modifying this landscape to destabilize these local minima and push the system into an avalanche.  That these avalanches display DRLO, is a feature of the added ``rigidity terms" in Eq.~(\ref{vdotrigid}). 

\section{Dynamical Long-Range Order}\label{DLROsec}

The discussion of DLRO in continuous dynamical systems is complicated by the continuity of the dynamics, making it difficult to clearly infer causal relationships between changes in variables.  However, we can take advantage of the timescales above to separate the dynamics into causally related events.  As shown in Fig.~\ref{fig:trajectories}(a) when we slow the memristor timescale $\beta$ relative to that of the voltages (e.g, by choosing $\beta=1/400$), after the initial transient the dynamics progress through a series of rapid transitions interpretable as \emph{avalanches} (or instantons in the field-theory language~\cite{topo,topo1}).  After an initial relaxation in which the gradient dynamics of the voltages rapidly seek out critical points in the energy landscape,  the slow evolution of the memristive dynamics transforms the stability of these points~\cite{DMMperspective,topo,topo1}, leading to a subsequent relaxation.   As more constraints become satisfied and transition to a rigid interaction, larger clusters of voltages begin transitioning together (see Fig.~\ref{fig:trajectories}(c)) in a manner analogous to the dynamics of physical systems in the vicinity of a phase transition~\cite{DMMperspective,sheldon2017}.  We note that in contrast to gradient dynamics the energy is {\it not} guaranteed to decrease monotonically, but the transitions induced by the memristive variables allow the system to reach lower values than are achievable through gradient dynamics alone.

We now argue that the existence of DLRO in the dynamics of the memcomputing solver is strongly connected to its ability to solve an optimization instance.  As a measure of DLRO we compute correlation functions over the largest avalanche that occurs in a simulation.  In the limit that the timescales become separated (i.e., the slow driving limit) the points at which each avalanche occurs tend towards well defined times as seen in Fig.~\ref{fig:trajectories}(a). For small $\beta$ these events may be detected as  sharp spikes in the voltage derivatives. (See Supplemental Material for a detailed discussion of the method used to extract the structure of the avalanches.) We are interested in the voltages/spins which change sign in the avalanche and thus will affect the energy of the system. We define the avalanche configuration as $\Delta_{i} = 1$ for all spins which change sign during an avalanche, and $\Delta_{i}=0$ otherwise.  A few typical examples of these avalanches and their sizes occurring during dynamics are plotted in Fig.~\ref{fig:trajectories}(c).  

\begin{figure}[t!]
	\begin{center}
		\includegraphics[width=8.6cm]{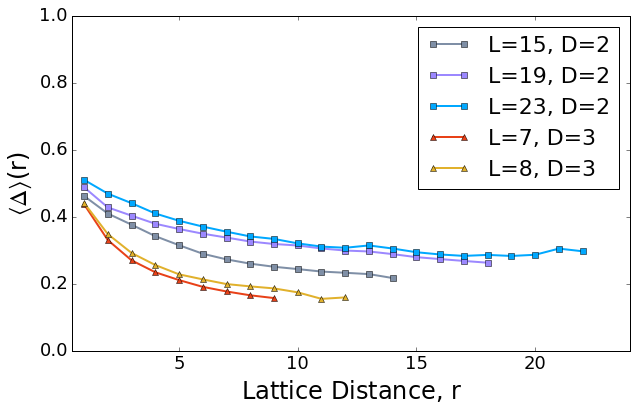}
		\caption{ Spatial correlations, $\langle \Delta\rangle(r)$, among voltages/spins calculated from the orthant dynamics in the slow driving limit 
			of model~(\ref{vdotrigid}) and~(\ref{xdotrigid}) for the largest avalanches in 2D and 3D and for different lattice sizes.  The correlations take a finite value all the way to the lattice edge, indicating that the largest avalanches are extensive.  As the system size increases the values appear to saturate to a dimension-dependent value for this instance class.
		\label{fig:correlations}}
	\end{center}
\end{figure}

Using the avalanche configurations we are able to compute correlation functions for these events and investigate their decay across the lattice. For each run (defined as generating a unique instance and initial conditions) the system is simulated until it reaches the ground state or a maximum time is reached. If the instance is solved within this interval,  the largest avalanche is selected and its configuration and first flipping spin are stored.   By averaging across a sample of configurations generated on different instances and initial conditions, suitably shifted so that the initial flipping spins coincide, the probability that a voltage a distance $r$ from the initial spin changes sign, $\langle \Delta\rangle(r)$, may then be calculated.  In order to achieve large distances with reasonable simulation times, we calculated these correlations both in 2- ($L=15, 19, 23$) as well as 3-dimensional ($L=7, 8$) systems.  For the parameters tested, success probabilities ranged from 92.8\% to 98.6\% depending on size and dimension.

As shown in Fig.~\ref{fig:correlations}, the largest avalanches possess correlations that take finite values all the way to the furthest corner of the lattice, manifesting a form of DLRO. Dimensionally, this requires that the size of the largest avalanche scales as $\sim L^D$ for a system of dimension $D$, and is thus extensive.  We also note that, as the system size increases the correlations appear to saturate to a dimension (and instance class) dependent value.

\begin{figure}[t!]
	\begin{center}
		\includegraphics[width=8.6cm]{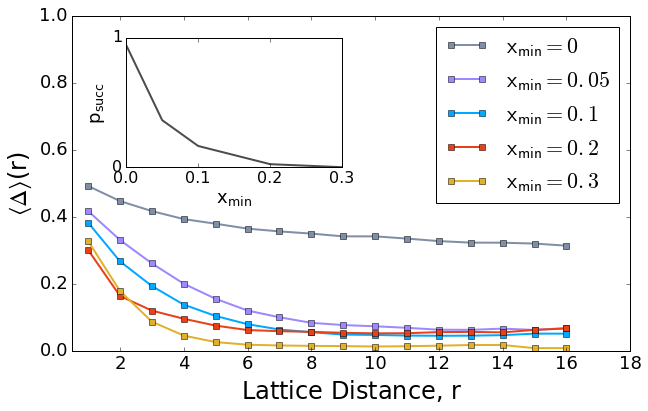}
		\caption{ Correlations and success probability (inset) as the minimum memristive value $x_{min}$ is varied. When $x_{min}=0$, no barrier remains between the two satisfied states of the interaction and the avalanche configurations are most long-ranged.  As $x_{min}$ increases, both the correlations and the success probability decay rapidly such that by $x_{min} = 0.3$ no instances are solved.
		\label{fig:xmin}}
	\end{center}
\end{figure}

We can control the presence of DLRO by adjusting the relative magnitudes of terms in the equations of motion.  First, we set a limit to the minimum size of $x_{ij}$ such that the barrier between the two satisfied states does not completely vanish.  This is accomplished by changing the memristive equation $\dot{x}_{ij}$ to,
\begin{equation}\label{eqn:xmin}
\dot{x}_{ij} = \beta (x_{ij} - x_{min})(1-x_{ij})\Big(|J_{ij}|\big(1-\text{sgn}(J_{ij}) v_i v_j\big) - \gamma\Big).
\end{equation}
This is simulated for $L=16,$ $D=2$, $\beta = 1/518$ and $x_{min} = 0,\, 0.05,\, 0.1,\, 0.2,\,0.3$.  At each value of $x_{min}$ correlations and success probabilities were calculated by averaging over 1000, 1000, 800, 600, and 500 runs, respectively, with the results displayed in Fig.~\ref{fig:xmin}.  Runs at larger values of $x_{min}$ are more computationally expensive due to the low success probability.  Further details on the simulations are included in the Supplemental Material. 

We observe that both the correlations and the success probability decay rapidly as $x_{min}$ is moved away from zero such that by $x_{min} = 0.3$ the success probability has vanished.  In a continuous dynamical system, the lack of energy barriers as $x_{min}\to 0$ is directly connected to the emergence of `zero-modes' along which the system can show collective behavior. We also note that as gradient dynamics correspond to $x_{min} = 1$, none of the instances tested would be solved by gradient dynamics alone suggesting the existence of many spurious critical points in this instance class.

\begin{figure}[t!]
	\begin{center}
		\includegraphics[width=8.6cm]{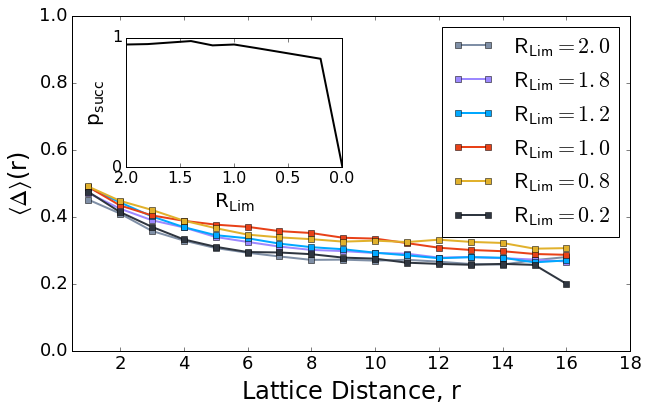}
		\caption{ Correlations and success probability (inset) as the magnitude of the rigidity terms are varied. In contrast to varying $x_{min}$ (as in Fig.~\ref{fig:xmin}) which introduces a barrier between the two satisfied states, adjusting the magnitude of $R_{Lim}$ maintains the ability of interacting variables to freely transition between their satisfied states.  We observe a wide range in which $R_{Lim}$ has a negligible effect both on the magnitude of avalanche correlations and the success probability.
		\label{fig:RLim}}
	\end{center}
\end{figure}

We can also vary the magnitude of the added `rigidity terms' independently, modifying our equation of motion of the voltages (Eq.~(\ref{vdotrigid})) to,
 \begin{align}
 \dot{v}_i  =& \sum_{\langle ij\rangle} J_{ij} x_{ij} v_j - \nonumber \\
 &\quad R_{Lim}(1-x_{ij})\frac{|J_{ij}|}{2}(v_i - \text{sgn}(J_{ij})v_j), \quad v_i \in [-1, 1].\label{eqn:R_{Lim}} 
 \end{align}
 These were simulated on the same instance class for $R_{Lim}$ values from 0 to 2, averaging across 300 runs at each value, with the results shown in Fig.~\ref{fig:RLim}.   In this case the ability of variables to transition collectively is maintained and in contrast to the previous results, we observe a wide range in which varying the size of $R_{Lim}$ has almost no effect on both the correlations and the success probability, and that the success probability vanishes as the size of the rigidity terms is brought to 0.  We interpret this as evidence that during the dynamics, voltages whose interaction has been satisfied remain close to the minimum of the interaction term (see Fig.~\ref{fig:interaction}(b)) and as such are not sensitive to the overall magnitude of this term.  Correlations are not plotted for $R_{Lim} = 0$ as these do not undergo qualitatively similar dynamics and the avalanche detection scheme described in the Supplemental Material fails.  
 
\section{Conclusions}\label{conclusions}

In this paper, using the frustrated-loop instances based on the Ising spin glass as a well-known benchmark, we have shown that a solver exploiting dynamical long-range order can navigate a non-convex landscape more efficiently than traditional methods based on annealing despite being composed of only local connections.   First, using a full implementation of DMMs we have shown results on 3D frustrated loop instances which indicate this approach is extremely effective in converging to the ground state solution.  In particular, DMMs demonstrated polynomial scaling in reaching the ground state on the tested instances, while all other solvers we have employed scaled exponentially, or possibly polynomially but with a substantially larger degree.

As the effectiveness of DMMs has been attributed to the presence of dynamical long-range order (DLRO) in the dynamics, we have constructed a simple model based on the structure of DMMs in which long-range behavior was introduced heuristically, allowing us to probe the connection between the presence of DLRO and the success probability of the solver.  By calculating correlations (or flipping probabilities) on the largest avalanche in the dynamics, we have demonstrated the existence of a form of long-range order which allows a transition to flip spins spanning the entire lattice.  By varying the parameters of the model, we have further shown that the magnitude of these correlations and the success probability are controlled by the presence of zero-modes along which variables can respond in a correlated manner.  The utilization of this effect is enabled by the use of \emph{continuous} dynamical systems to solve natively discrete problems. The results presented here further reinforce the advantages of employing collective dynamics to compute hard problems efficiently.

{\it Acknowledgments --} F.S. and M.D. acknowledge partial support from the Center for Memory and Recording Research at UCSD. The Falcon solver used in the 
reported simulations has been provided by MemComputing, Inc. \url{http://memcpu.com/}.
The authors would be delighted to provide, upon request, all instances of the spin-glass problems used in this work.

\newpage

\section{Supplemental Material}
\subsection{Correspondence between statistical physics and combinatorial optimization}\label{opt}

The problem of finding the ground state of a system in statistical physics is an optimization problem for which there is an extensive vocabulary in computer science~\cite{MooreMertens_book}.  It may be useful for the reader familiar with physics to have some notion of this correspondence and we include a short discussion here to that effect.

Minimizing a Hamiltonian expressed as a sum over interactions between spins may be represented as a constraint satisfaction problem where each term in the Hamiltonian is regarded as a constraint on the variables.  For example, finding the ground state of the Ising Model
\begin{equation}\label{eqn:Ising}
E = -\sum_{\langle ij\rangle} J_{ij} s_i s_j , \;\; s_i \in \{-1, 1\},
\end{equation}
is equivalent to a weighted constraint satisfaction problem, where each interaction is expressed as an exclusive-OR (XOR), or sum modulo-2 between the associated binary variables, $b_i = (s_i + 1)/2$,  $j_{ij} = (1-\text{sgn}(J_{ij}))/2$. We specify the constraint satisfaction problem by a list of weights and constraints given by the correspondence,
\begin{equation}\label{eqn:IsingXOR}
-J_{ij} s_i s_j \quad\leftrightarrow\quad 2|J_{ij}|\; \;b_i \oplus b_j = j_{ij}
\end{equation}
where $2|J_{ij}|$ is the weight associated with violating the constraint $b_i \oplus b_j = j_{ij}$. The correspondence between these two can also be easily seen from the fact that flipping the state of any variable changes the state of the interaction in both cases.  Transformations between other constraint satisfaction problem types may be undertaken similarly.  For example, when transforming to weighted conjunctive normal form (CNF), each interaction may be translated to two OR constraints depending on the sign of the interaction:
\begin{equation}\label{eqn:IsingOR}
-J_{ij} s_i s_j \quad\leftrightarrow\quad 
\begin{cases}
\substack{ 2|J_{ij}| \;b_i \vee \bar{b}_j \\ 2|J_{ij}| \;\bar{b}_i \vee b_j }, & \text{sgn}(J_{ij}) =1,   \\[8pt]
\substack{ 2|J_{ij}| \;b_i \vee b_j \\ 2|J_{ij}| \;\bar{b}_i \vee \bar{b}_j }, & \text{sgn}(J_{ij}) =-1, 
\end{cases}
\end{equation}
where each constraint carries a weight and negations are indicated with a bar, e.g., $\bar{b}$.  In all cases, the factor of 2 may be dropped as a global scaling of the energy.

Constraint satisfaction \emph{instances} (a particular example of the problem) may be described as being either satisfiable (SAT), if there is an assignment of the variables which satisfies every constraint, or unsatisfiable (UNSAT) if there is no satisfying assignment (commonly referred to as \emph{frustrated} in physical treatments). 

The corresponding \emph{decision problem} of determining whether such an assignment exists, and, therefore, whether a particular instance is SAT or UNSAT is also referred to as SAT or satisfiablity with context generally determining which meaning is intended.  The \emph{optimization problem} of determining an assignment which satisfies the maximum number of constraints (or maximum total weight) is referred to as MAXSAT.  Determining the ground state of a system in statistical physics is thus equivalent to a MAXSAT instance.

Generally, the SAT problem on 2-variable OR and XOR constraints may be trivially solved.  In the case of the Ising model, pick the value of any spin to be $+1$ and propagate this throughout the lattice where every spin value will be determined by its neighbor.  If a contradiction is reached, the instance is unsatisfiable.  If not, this will construct a satisfying assignment and the instance is equivalent to the ferromagnetic Ising model through a gauge transformation.

Despite this, the  MAXSAT problem on two variable constraints may be quite difficult, depending on the structure of the instance.  It is known that instances on a planar graph may be solved efficiently (in polynomial time) by a perfect matching algorithm~\cite{mandra2017pitfalls}. If the graph is non-planar as in the chimera graphs used by DWave~\cite{DenchevGoogle} or the 3-dimensional cubic lattice used for benchmarking here, there is no general efficient algorithm known, and the problem of finding an assignment is NP-Hard~\cite{barahona1982computational}.  This statement, however, only applies to the worst cases and for any individual instance, and especially for classes of randomly generated instances, one might hope that an efficient approach exists.  Conversely, despite the fact that an efficient algorithm exists for planar instances, they may still present meaningful difficulty for a solver which only uses local information.  Debate over these ideas have surrounded the benchmarking studies for DWave and discussions to this effect may be found in~\cite{Katzgraber2015,mandra2017pitfalls,king2015performance,hen2015probing}.

\subsection{Avalanche Detection and Correlation Calculations}

\begin{figure}
	\begin{center}
		\includegraphics[width=8.6cm]{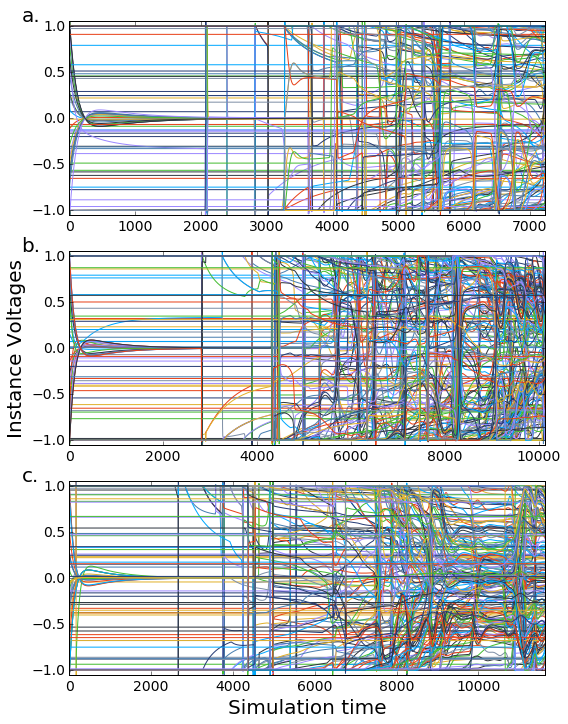}
		\caption{Here we show the simulation results for the same initial conditions on three different instances ($L=15,\,D=2,\,\alpha=0.2$).  In (a.) we see well separated events that maintain their separation until the instance is solved.  In (b.) the longer run results in lower memristor values and a slower voltage timescale, causing the width of the avalanches to grow.  In (c.) we see markers of quasi-periodic behavior extending from $\approx 8000$ to $10000$.  A scheme to detect avalanches must be robust to these effects in order to be accurate.
		\label{fig:trajs}}
	\end{center}
\end{figure}

As shown in the main text, when the timescale of the memristors is sufficiently separated from the voltage timescale, the system evolves in well defined events which may be interpreted as avalanches (or instantons~\cite{topo}). A set of sample trajectories, calculated on different spin glass instances ($L=15,\,D=2,\,\alpha=0.2$) in the slow driving limit is displayed in Fig.~\ref{fig:trajs}. Within these trajectories avalanches are clearly discernible, but the timescales governing them seem to grow during the simulation and additional features such as quasi-periodic behavior (as seen in panel~\ref{fig:trajs}.c  at beginning time 8000) also arise. Detecting these events in a robust manner poses a novel problem as we must extract a discrete event from a continuous system.  Here we detail the method we used to extract these events and compute correlations during the largest avalanche.

\begin{figure*}
	\begin{center}
		\includegraphics[width=17.2cm]{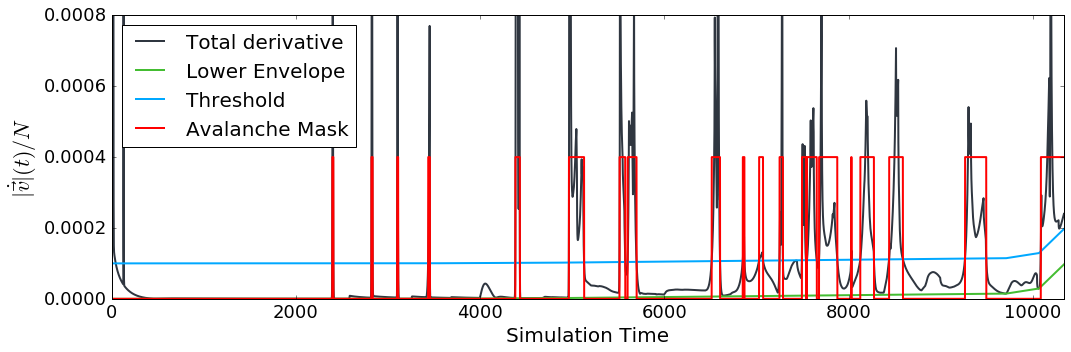}
		\caption{ Avalanches display as sharp spikes in the total voltage derivative 
			vector of the system.  Here, we have displayed the magnitude of the total voltage derivative of the system normalized to the number of spins N for the trajectory displayed in the main text ($L=15,\, D=2\, \alpha=0.2$).  As the timescale of the avalanches can slow, sometimes dramatically, over the course of the simulation, the convex lower bound of the derivative is first calculated.  An avalanche interval is defined as a continuous  period in which the system exceeds a threshold above this envelope (here chosen as 0.0001).  Voltages that change sign during an interval are included in the avalanche configuration as shown in Fig.~\ref{fig:clusters}.
		\label{fig:deriv}}
	\end{center}
\end{figure*}

The rate of change of the entire system may be concisely viewed through the magnitude of the voltage derivative vector, $\dot {\vec{v}}(t)$, normalized to the number of spins,  $|\dot{\vec{v}}(t)|/N$ shown in Fig.~\ref{fig:deriv}.    Avalanches manifest as sharp spikes in the magnitude of the derivative. However, as the simulation advances and variables begin transitioning together, the slowest timescale in the system tends to increase, making a simple threshold ineffective at separating the later clusters.

\begin{figure*}
	\begin{center}
		\includegraphics[width=17.2cm]{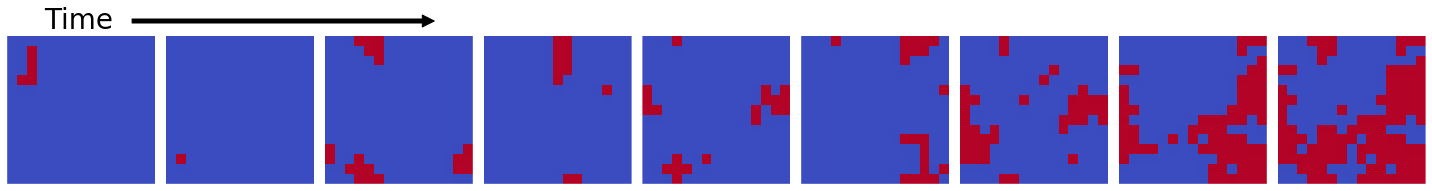}
		\caption{ For the trajectory shown in the main text and the clusters detected in Fig.~\ref{fig:deriv} a representative sample of the detected avalanche configurations are displayed, including the first and last avalanches in the trajectory.  Over the course of the simulation, the average size of the avalanches grows until it reaches an extensive set of spins which can span the entire lattice.
		\label{fig:clusters}}
	\end{center}
\end{figure*}

Instead, we first find the convex lower envelop of the total derivative shown in Fig.~\ref{fig:deriv}.  This gives an estimate of how the slowest timescale in the system changes.  If the slope at the end of the envelope exceeds a bound, its slope is extrapolated from the previous point to avoid errors due to the termination of the integration mid-avalanche.

The time interval of an avalanche is defined as a continuous period in which the magnitude of the derivative $|\vec{\dot{v}}(t)|/N$ exceeds the lower envelop by a threshold.  Choosing this threshold is performed through tuning to the specific set of instances and will depend on system size, dimension and memristor timescale $\beta$.  The threshold value $t$ used for each set of correlation calculations are: $(L=15,\,D=2,\,t=1\times 10^{-4})$, $(L=19,\,D=2,\,t=2.15\times 10^{-5})$, $(L=23,\,D=2,\,t=8\times 10^{-6})$, and $(L=8,\,D=3,\,t=4\times 10^{-6})$ .  Within an avalanche interval, we define the variables included in the avalanche as those that changed sign and thus can affect the energy calculated from the orthant dynamics. A few of these configurations are shown in the main text and a more complete selection is displayed in Fig.~\ref{fig:clusters}.

Once a set of avalanches has been extracted from a simulation, we investigate the structure of the largest avalanche by calculating the probability that a spin lying a distance $r$ away (measured in terms of lattice steps) from the first spin to flip is included within the avalanche.  To this end, we define a cluster configuration as being $v_i = 1$ if the spin is included in the cluster and $0$ otherwise.  This acts as indicator variable which for an individual cluster allow us to calculate the probability that a spin a distance $r$ away was flipped (recall that the lattices we generate are periodic and this distance is calculated as the minimum distance of a path between the two sites).  This is then averaged across randomly generated instances and initial conditions of the solver.

The probability obtained may be interpreted as a correlation in the slow driving or instantonic limit in which the avalanche may be regarded as occurring at an instant in time, and calculated on the orthant dynamics of the system.  As shown in the main text, the largest avalanche gives a finite probability for a spin anywhere in the lattice to change sign and is thus extensive.

\subsection{Simulations and Solver Tuning}

With the exception of IBM CPlex and Falcon, solvers were implemented in Python 2.7 using the NumPy and SciPy libraries~\cite{scipy}.  Simulated Annealing, Parallel Tempering, CPlex and the model in the main text were run at UCSD on a single core of an Intel Xeon E5430 with 16 Gb RAM with a peak rate of $2\,\text{operations/cycle} \times 2.66\, \text{GHz} = 5.32 \, \text{Gflops}$.  The time to solution in seconds of the solvers is displayed in Figure~\ref{fig:scaling}.

\begin{figure}[t]
	\begin{center}
		\includegraphics[width=8.6cm]{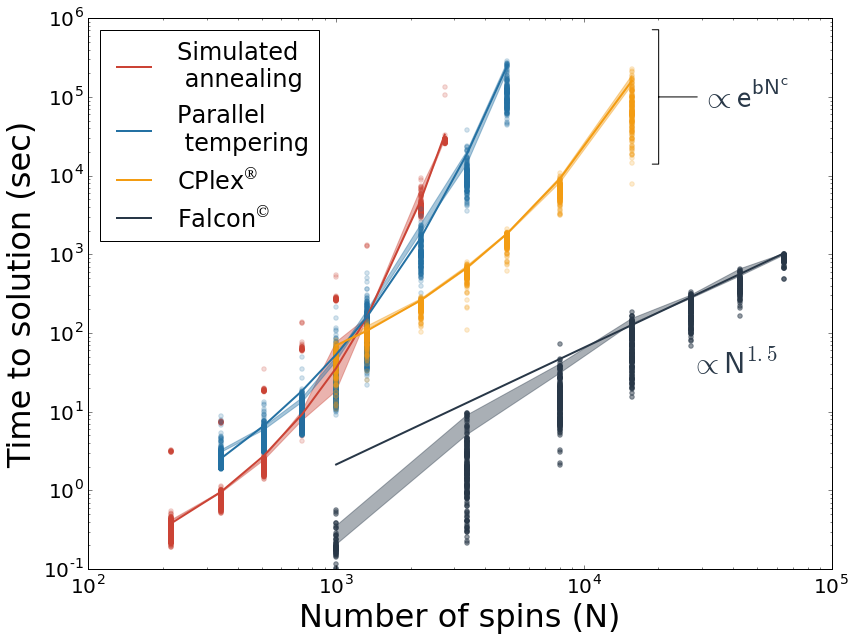}
		\caption{ Time necessary for solvers to reach the ground state of the 3D frustrated-loop spin glass as a function of the total 
			number of spins $N$. The sequential memcomputing solver implemented in MATLAB is dubbed Falcon. Varying numbers of instances were run at each size and solver depending on required computation time (See the Supplemental Material for details). Comparisons with simulated annealing (SA), parallel tempering (PT), and IBM CPlex are also shown. All calculations were performed on a single core. The solid lines are the best fits of the 95th quantile time to solution for all four solvers. The exponential fits have the following parameters: for IBM CPlex, $b=0.12$ and $c=0.46$, for SA, $b=0.069$ and $c=0.67$, and for PT $b=0.32$ and $c=0.46$.}\label{fig:scaling}
	\end{center}
\end{figure}

\subsubsection{Model Simulations}

In order to limit the voltages and memristors to the allowed regions and make them robust to numerical errors, the equations simulated were,
\begin{align}\label{eqn:numerical}
\dot{v}_i & = B_{v_i, (-1,1)}\bigg( \sum_{\langle ij\rangle} J_{ij}x_{ij}v_j \nonumber \\
& \qquad {}-(1-x_{ij})\frac{|J_{ij}|}{2}\big(v_i - \mathrm{sgn}(J_{ij}) v_j\big) \bigg) \\
\dot{x}_{ij} & = \beta B_{x_{ij}, (0, 1)} \bigg( x_{ij} \big(1-x_{ij}\big) \nonumber \\
& \qquad\times\Big(\frac{|J_{ij}|}{2}\big(1-\mathrm{sgn}(J_{ij})v_i v_j \big) - \gamma\Big) \bigg)
\end{align}
where $B_{x, (l, h)}(\cdot)$ implements the bounds to ensure integration steps that leave the region return to the fixed points as
\begin{equation}\label{eqn:bounding}
B_{x, (l, u)}(f) = \begin{cases}
(u-x), & x > u, f > 0 \\
(l-x), & x < l, f < 0 \\
f, & \text{otherwise.}
\end{cases}
\end{equation}
Integrations are carried out using forward Euler with parameters tuned for each set of instances to maintain the slow driving limit, hence allowing for 
an easy identification of avalanches.  This tuning is not required to \emph{solve} an instance, but it is in order to detect clearly defined avalanches.  First, it was determined through tuning that $\beta=\frac{1}{400}$ with a maximum time of $t_{max} = 25,000$ gave well defined avalanches for $L=15,\,D=2$ and usually solved instances near to $t=t_{max}/2$.  In order to maintain this limit for larger instances, the memristor timescale was slowed, scaling with the inverse square of the number of spins.  Slowing the memristor timescale requires increasing the maximum simulation time in the same way, such that $t_{max}$ was scaled with the square of the number of spins.   For $L=15$, $\gamma=0.65$ was found to give a well defined transition, but as instance size increased this value would lead to quasi-periodic behavior more often and $\gamma$ was increased to $L=19,\,\gamma=0.75$, $L=23,\,\gamma=0.85$ and for $L=7,\, D=3$ and $L=8,\,D=3,\,\gamma=0.85$.

Simulations varying $x_{min}$ and $R_{Lim}$ were carried out on an $L=16$ square lattice in $D=2$ with $\gamma=0.7$ and $\beta$ and $t_{max}$ scaled as described above.

\subsubsection{Simulated Annealing}

\begin{figure}[t!]
	\begin{center}
		\includegraphics[width=8.6cm]{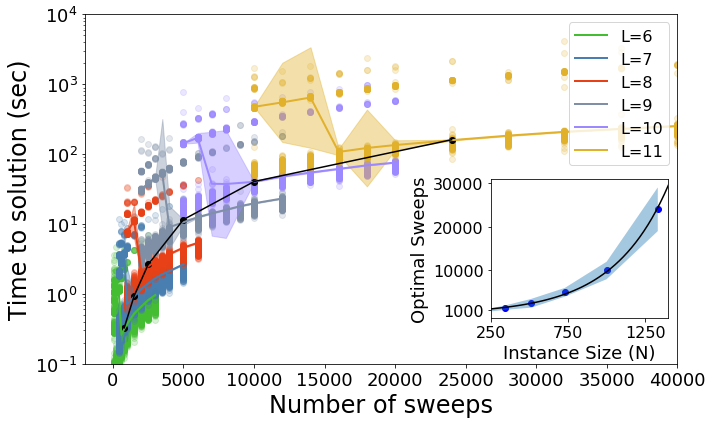}
		\caption{Simulated annealing runs were performed on 1000 3-D frustrated loop instances for $L=6, 7, 8 ,9, 10, 11$ at varying number of metropolis sweeps.  As can be seen, if the number of sweeps is too few the ground state is only rarely encountered.  Beyond a certain amount, more sweeps will have a  slightly negative effect on the runtime.  We estimated the location of this crossover across lattice sizes with the dark curve, and extracted the scaling of the optimal runtime (inset) which was well fit by an exponential.
		\label{fig:SAtuning}}
	\end{center}
\end{figure}

Simulated annealing was implemented using a linear schedule in $\tilde \beta$ (the inverse temperature) from $\tilde \beta_i = 0.01$ to $\tilde \beta_f=\ln (N)$ where the low temperature was scaled such that excited states were suppressed as $\frac{1}{N}$~\cite{white1984concepts}.  At each step in the annealing a sweep of metropolis samples across the entire lattice was performed.  For annealers to be most effective in a given computational time, the number of cooling steps (i.e., number of metropolis sweeps) performed on a single initial condition versus the number of initial conditions attempted must be optimized.  As we are working with planted solution instances, runs consisted of continued repetitions at a fixed number of metropolis sweeps until the solution energy was encountered or some maximum allowed time was reached.  Tuning was performed by running each instance with a varying number of temperature steps in the cooling schedule until the optimum was reached.  This was performed on 1000 frustrated loop instances ($d=3$, $\alpha=0.3$, $l\ge 6$) for $L=6$ through $11$ and the time to solution for each run was recorded.  The results of these runs are plotted in Fig.~\ref{fig:SAtuning}.

The time to solution for a given instance size and number of sweeps exhibited a bimodal distribution with a secondary peak of smaller measure approximately one order of magnitude above the primary peak.   In order to estimate the optimal scaling, the 95th quantile of each sample distribution was found and the error estimated with a bootstrapping procedure (30,000 samples).  As the peaks of the distributions were approximately lognormal, these were both calculated in logspace and the error estimated as $2\sigma$ of the sample distribution.  As can be seen in Fig.~\ref{fig:SAtuning}, the presence of a secondary peak leads to large uncertainties in the region surrounding the cross-over of the 95th quantile.  The optimal sweeps were found as the minimum of the upper bound uncertainty estimate.

Since the intention of this work is to examine the scaling for very large sizes, this analysis cannot be repeated for all sizes we intended to run.  Instead, we used the values at small $N$ to estimate the scaling of the optimal number of sweeps per repetition as shown by the inset curve in Fig.~\ref{fig:SAtuning}.  This was well fit by an exponential dependence as,
\begin{equation}\label{optsweeps}
\# sweeps_{opt} = (604) \exp (2.78\times 10^{-3} N)
\end{equation}
which was used to estimate the optimal number of sweeps for the scaling figure in the main text.

\begin{figure}[t!]
	\begin{center}
		\includegraphics[width=8.6cm]{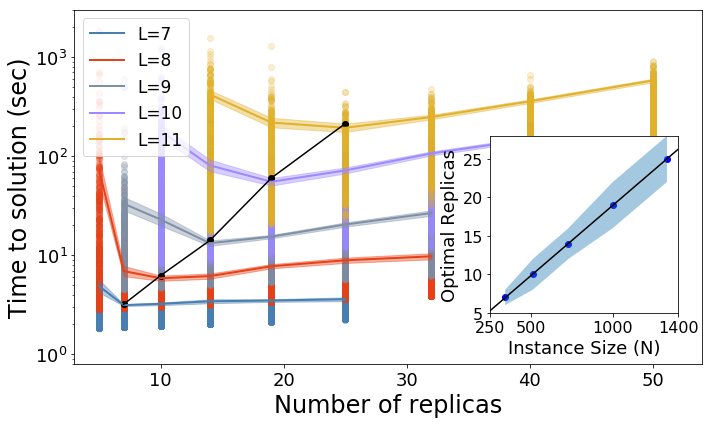}
		\caption{ Parallel tempering runs were performed on 1000 frustrated loop instances for $L=6, 7, 8, 9, 10, 11$ with varying numbers of replicas.  For sufficiently large sizes ($L>6$) a clear optimum was observed in the time to solution, with the number of replicas at the optimum growing with the number of spins.  The location of these optima were used to estimate a scaling law for the optimal number of replicas at larger sizes.
			\label{fig:PTtuning}}
	\end{center}
\end{figure}

\subsubsection{Parallel Tempering}

The parallel tempering (PT) algorithm we employed utilized a ladder of temperatures geometrically spaced in $T$ from $T_h = 10$ to $T_l = 1/\log N$~\cite{white1984concepts,earl2005parallel}.  One step of the algorithm consisted of a single sweep of metropolis sampling over all replicas, followed by a single proposed exchange where replicas at neighboring temperatures had their configurations switched according to the probability,
\begin{equation}\label{eqn:exchange}
P_{exchange} = \min\{1, \exp\Big((\tilde \beta-\tilde \beta')(E-E') \Big) \},
\end{equation}
where $E$ and $E'$ are energies of the instances. 
This cycle of metropolis sweeps and exchanges was repeated until the solution was reached.  As the system size increases, the extensivity of the energy will cause exchanges to become less likely and so the density of temperatures simulated should be increased such that the optimal temperature set will balance the extra computational work of performing metropolis sweeps over the replicas with the diffusion of replicas across temperatures.  In order to determine this point, the time to solution was found for 1000 frustrated loop instances (dimensions $D=3$, density $\alpha=0.3$, loop length $l\ge 6$) for $L=6$ through $11$ across a range of different numbers of replicas, as shown in Fig.~\ref{fig:PTtuning}. At each number of replicas and size, the 95th quantile of the sample was found and the error in this statistic was estimated from a bootstrap in logspace using a $2\sigma$ confidence interval.  The optimal number of replicas was found as the minimum of the upper bound error estimate.  Across sizes these were found to be very well fit by a linear dependence,
\begin{equation}\label{optsweeps}
\# replicas_{opt} = 0.0183 N + 0.7.
\end{equation}
This leads to a quadratic scaling in the memory footprint of the algorithm.

\subsubsection{CPlex}

CPlex was run using the python API within the IBM ILOG CPLEX Optimization Studio version 12.7.1.0 under an academic license~\cite{cplex}.  The QUBO (quadratic unconstrained binary optimization) form for the associated frustrated loop instance was found through the transformation to binary variables, $s_i = 2b_i-1$ which leads to the correspondence,
\begin{eqnarray}\label{eqn:SG2QUBO}
E_{SG} &= -\frac{1}{2}\sum_{ij}J_{ij} s_i s_j \\
{}& = \frac{1}{2}\sum_{ij}Q_{ij}b_i b_j + C \\
Q_{ij} &= \begin{cases}
-4J_{ij}, & i\neq j \\
4\sum_j J_{ij}, & i = j \end{cases} \\
C &= -4\sum_{ij} J_{ij}.
\end{eqnarray}
Within CPlex, problems in this form are first transformed to a mixed integer programming (MIP) form.  Unlike the other solvers in this work, CPlex is a complete solver and will attempt a proof of optimality along with solving the instance.  To prevent this, a callback was employed that terminates the search once the planted solution energy was found.  Cuts were set to balance optimality and feasibility in the search. 

\subsubsection{Falcon}

The memcomputing solver, Falcon, was implemented with MATLAB as specified in~\cite{DMM2} as a Boolean satisfiability solver which accepts instances in conjunctive normal form~\cite{complexity_bible}.  The transformation to CNF shown in section I has been performed on the frustrated loop instances and simulations were carried out on a single core of an Intel Xeon 6148 with 192 GB RAM with a peak rate of $2\,\text{operations/cycle} \times 3.7\, \text{GHz} = 7.4 \, \text{Gflops}$. The Falcon parameters were first tuned for the 
smallest size instances, and then the same parameters have been employed to solve all instances reported in Fig. 4 of the main text. Since Falcon integrates differential equations numerically (using forward Euler), it employs memory that scales 
linearly with problem size~\cite{stress-testing}. 

\subsection{Fitting}

\begin{figure*}[t!]
	\begin{center}
		\includegraphics[width=17.2cm]{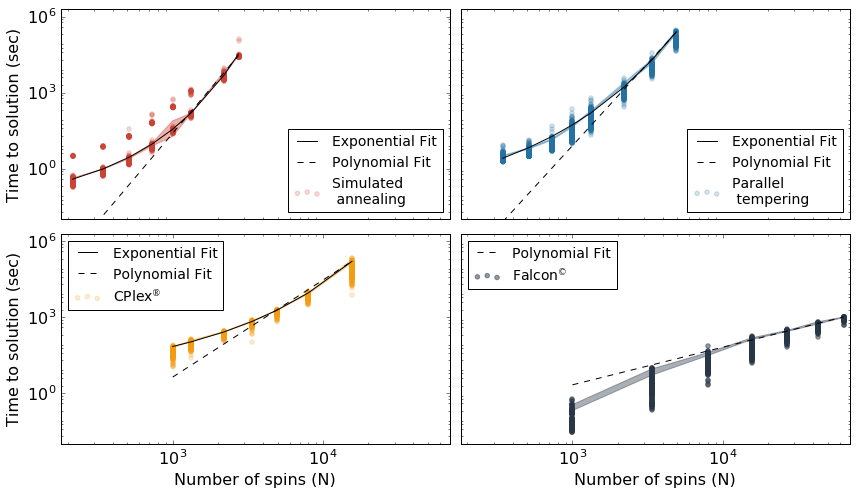}
		\caption{ For each solver, the geometric average complexity of the instance class was estimated by fitting the 95th quantile time to solution for a set of frustrated loop spin glasses ranging in sizes across several orders of magnitude.  Polynomial $(TTS\approx A N^B)$ and exponential $(TTS\approx a e^{bN^c})$ trends were fit to the data to assess the performance of each approach.  For Falcon, only the polynomial fit was found to converge and it demonstrated the lowest exponent of all fits ($B=1.5$).  All other solvers were found to scale exponentially.
			\label{fig:fits}}
	\end{center}
\end{figure*}

For each solver, instances were run at a variety of sizes in order to estimate the scaling of the 95th quantile time to solution.  The annealers, SA and PT were run on 1000 instances for lattices $L\le 11$ and 200 for $L>11$.  CPlex was run on 200 instances from $L=10$ to $L=25$ and Falcon was run on 200 instances for sizes from $L=10$ to $L=30$ and 100 for $L=35, 40$ (all in 3-dimensions).  The error in the sample 95th quantile estimate was calculated with a $2\sigma$ estimate from a bootstrap (30,000 samples) performed in log space to account for the lognormal distribution of the runtimes.  

Fitting of each solver was performed with the SciPy optimization library's curve fit and polyfit functions~\cite{scipy}, which utilize least squares to fit a particular function.  The time to solution for instances was found to be approximately log-normally distributed, as may be observed from the consistent standard deviations (0.5-1.5 orders of magnitude) found in log-space in Fig.~\ref{fig:fits}.  Fitting was thus performed in log-space on the sample 95th quantile time to solution.  For each solver, fits were performed for a polynomial, $TTS\approx A N^B$ and exponential $TTS\approx a e^{bN^c}$ trend as is displayed in Fig.~\ref{fig:fits}.  For Falcon, an exponential fit did not converge and so only a polynomial fit is shown following $TTS\approx (7.5\times 10^{-5}\, sec) N^{1.5}$.

For the other solvers, both a polynomial and exponential fit was found to converge but in each case the exponential fit is clearly favored.  For SA, the time to solution was found to follow $TTS\approx (0.030\,\text{sec})e^{0.069N^{0.67}}$, for PT $TTS\approx (0.022\,\text{sec})e^{0.32N^{0.46}}$, and for CPlex $TTS\approx (3.5\,\text{sec})e^{0.12N^{0.46}}$. It is interesting to note that the exponential coefficients for each solver are quite close to rational values, with SA scaling close to $e^{bN^{2/3}}$ and PT and CPlex scaling close to $e^{bN^{1/2}}$.

%

\end{document}